\documentclass[prl,amssymb,twocolumn,tightenlines,showpacs]{revtex4-1}

\usepackage{graphicx}
\usepackage{amsmath}
\usepackage{verbatim}
\usepackage{subfigure}

\begin{document}

\title{Entanglement Between Photons that have Never Coexisted}

\author{E. Megidish}
\affiliation{Racah Institute of Physics, Hebrew University of
Jerusalem, Jerusalem 91904, Israel}
\author{A. Halevy}
\affiliation{Racah Institute of Physics, Hebrew University of
Jerusalem, Jerusalem 91904, Israel}
\author{T. Shacham}
\affiliation{Racah Institute of Physics, Hebrew University of
Jerusalem, Jerusalem 91904, Israel}
\author{T. Dvir}
\affiliation{Racah Institute of Physics, Hebrew University of
Jerusalem, Jerusalem 91904, Israel}
\author{L. Dovrat}
\affiliation{Racah Institute of Physics, Hebrew University of
Jerusalem, Jerusalem 91904, Israel}
\author{H. S. Eisenberg}
\affiliation{Racah Institute of Physics, Hebrew University of
Jerusalem, Jerusalem 91904, Israel}


\begin{abstract}
The role of the timing and order of quantum measurements is not
just a fundamental question of quantum mechanics, but also a
puzzling one. Any part of a quantum system that has finished
evolving, can be measured immediately or saved for later, without
affecting the final results, regardless of the continued evolution
of the rest of the system. In addition, the non-locality of
quantum mechanics, as manifested by entanglement, does not apply
only to particles with spatial separation, but also with temporal
separation. Here we demonstrate these principles by generating and
fully characterizing an entangled pair of photons that never
coexisted. Using entanglement swapping between two temporally
separated photon pairs we entangle one photon from the first pair
with another photon from the second pair. The first photon was
detected even before the other was created. The observed quantum
correlations manifest the non-locality of quantum mechanics in
spacetime.
\end{abstract}

\maketitle

Entanglement between spatially separated quantum systems is one of
the most distinctive results of quantum mechanics. It results in
nonclassical correlations between distant systems. Einstein,
Podolsky, and Rosen claimed that these instantaneous correlations
give rise to a paradox which demonstrates the incompleteness of
quantum mechanics \cite{EPR}. Only after the realization of an
experiment suggested by Bell \cite{Bell,Aspect81}, was the
nonlocal nature of quantum mechanics widely accepted.
Nevertheless, the properties of entanglement still puzzle many
researchers.

Single photons are used as quantum particles in many experimental
realizations, as they are easily manipulated and preserve their
coherence for long times. A common method for generating
polarization entangled photon states is using the nonlinear
optical process of parametric down-conversion (PDC) in dielectric
crystals \cite{Kwiat95}. In this process, a pump photon splits
into two lower-energy photons while preserving momentum and
energy. With this method it is possible to create bright
high-quality two photon states in any of the four maximally
entangled states, also known as the Bell states. For polarized
photons these states are
\begin{eqnarray}\label{BellStates}
\nonumber &&|\phi^{\pm}\rangle=\frac{1}{\sqrt{2}}(|h_ah_b\rangle \pm |v_a v_b\rangle),\\
&&|\psi^{\pm}\rangle=\frac{1}{\sqrt{2}}(|h_a v_b\rangle \pm |v_ah_b\rangle)\,,
\end{eqnarray}
where $h_a (v_b)$ represents a horizontally (vertically) polarized
photon in spatial mode a (b).

Photons can also be entangled by projection measurements onto
maximally entangled states \cite{Weinfurter94}. Bell state
measurements with linear optical elements require post-selection.
They can discriminate simultaneously only between two of the four
Bell states \cite{Vaidman99,Lutkenhaus99}. Complete Bell
projection can be achieved using nonlinear optics \cite{Kim},
hyper entanglement \cite{Barreiro,Schuck}, auxiliary photons
\cite{Zhao05,Walther05,Grice11} or path entangled single photons
\cite{Boschi98}. Bell projections are key ingredients in quantum
computation and communication protocols such as teleportation
\cite{Bennett93} and entanglement swapping \cite{Zukowski93}.

The entanglement swapping protocol entangles two remote photons
without any interaction between them. Each of the two photons
belongs initially to one of two independent entangled photon pairs
(e.g., photons 1 and 4 of the entangled pairs 1-2 and 3-4). The
two other photons (2 and 3) are projected by a measurement onto a
Bell state. As a result, the first two photons (1 and 4) become
entangled even though they may be distant from each other.
Entanglement swapping is the central principle used in quantum
repeaters \cite{Briegel98}, whose purpose is to overcome the
limiting effect of photon loss in long range quantum
communication. Previous demonstrations of entanglement swapping
\cite{PanSwapping} and multi-stage entanglement swapping
\cite{PanSwapping6Photons}, entangled photons that were separated
spatially, but not temporally, i.e., all the photons that were
entangled, existed and were measured at the same time.

\begin{figure}[tb]
\centering\includegraphics[angle=0,width=86mm]{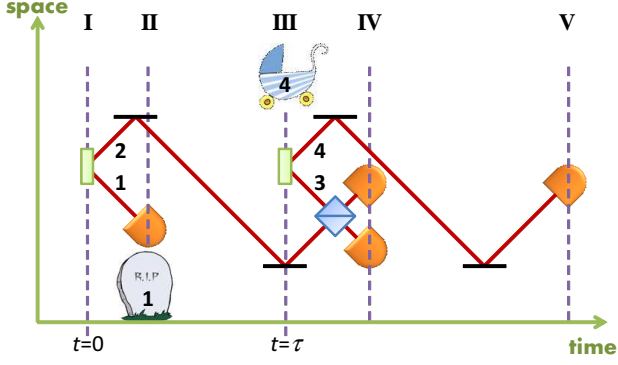}
\caption{\label{Storyboard}(color online). Time line diagram. (I)
birth of photons 1 and 2. (II) detection of photon 1. (III) birth
of photons 3 and 4. (IV) Bell projection of photons 2 and 3. (V)
detection of photon 4.}
\end{figure}

In this work we demonstrate how the time at which quantum
measurements are taken and their order, has no effect on the
outcome of a quantum mechanical experiment, by entangling two
photons that exist at separate times. This is achieved by first
creating one photon pair (1-2) and right away measuring photon 1
(see Fig. \ref{Storyboard}). Photon 2 is delayed until a second
pair (3-4) is created and photons 2 and 3 are projected onto the
Bell basis. When photon 1 is measured in a certain basis, it does
not 'know' that photon 4 is going to be created, and in which
basis it will be measured. Nevertheless, photons 1 and 4 exhibit
quantum correlations despite the fact that they never coexisted.
The possibility of two photons that do not overlap in time, but
still exhibit entanglement, was discussed theoretically in a
system of atoms and photons \cite{Wiegner11}. Recently,
entanglement swapping was demonstrated with a delayed choice,
where all four photons were created simultaneously, but photons 1
and 4 were measured before a choice had been made whether to
entangle them or not \cite{Peres00,Ma12}.

The scenario of time and space separation we create should be
compared to the standard two particle entangled state, where the
particles are only spatially separated. In the standard
entanglement case, the measurement of any one of the particles
instantaneously changes the physical description of the other.
This result was described by Einstein as "spooky action at a
distance". In the scenario we present here, measuring the last
photon affects the physical description of the first photon in the
past, before it has even been measured. Thus, the "spooky action"
is steering the system's past. Another point of view that one can
take is that the measurement of the first photon is immediately
steering the future physical description of the last photon. In
this case, the action is on the future of a part of the system
that has not yet been created.

In order to generate consecutive photon pairs at well defined
times, a pulsed laser is used to pump a single PDC polarization
entangled photon source \cite{Kwiat95}. It is a probabilistic
source, and thus there is a probability that two pairs will be
created, each pair from one of two consecutive pulses, separated
by the laser period time $\tau$. The four-photon state is
\begin{eqnarray}\label{InitialState}
\nonumber |\psi^{-}\rangle_{a,b}^{0,0}&\otimes&|\psi^{-}\rangle_{a,b}^{\tau,\tau}= \\
\frac{1}{2}(|h_a^{0}v_b^{0}\rangle -
|v_a^{0}h_b^{0}\rangle)&\otimes& (|h_a^{\tau}v_b^{\tau}\rangle -
|v_a^{\tau}h_b^{\tau}\rangle)\,,
\end{eqnarray}
where the subscripts are the spatial mode labels and the
superscripts are the time labels of the photons. In order to
project the second photon of the first pair and the first photon
of the second pair onto a Bell state, the former is delayed by
$\tau$ in a delay line. The same delay is also applied to the
second photon of the second pair and the resulting state can be
reordered and written as
\begin{eqnarray}\label{StateAfterDelay}
\nonumber
           |\psi^{-}\rangle_{a,b}^{0,\tau}\otimes|\psi^{-}\rangle_{a,b}^{\tau,2\tau}=
           \frac{1}{2}(
           &|&\psi^{+}\rangle_{a,b}^{0,2\tau}|\psi^{+}\rangle_{a,b}^{\tau,\tau}\\
\nonumber -&|&\psi^{-}\rangle_{a,b}^{0,2\tau}|\psi^{-}\rangle_{a,b}^{\tau,\tau}\\
\nonumber -&|&\phi^{+}\rangle_{a,b}^{0,2\tau}|\phi^{+}\rangle_{a,b}^{\tau,\tau}\\
+&|&\phi^{-}\rangle_{a,b}^{0,2\tau}|\phi^{-}\rangle_{a,b}^{\tau,\tau})\,.
\end{eqnarray}
When the two photons of time $\tau$ (photons 2 and 3) are
projected onto any Bell state, the first and last photons (1 and
4) collapse also into the same state and entanglement is swapped.
The first and last photons, that did not share between them any
correlations, become entangled.

According to this description, the timing of each photon is merely
an additional label to discriminate between the different photons,
and the time in which each photon is measured has no effect on the
final outcome. The first photon from the first pair (photon 1) is
measured even before the second pair is created (see Fig.
\ref{Storyboard}). After the creation of the second pair, the Bell
projection occurs and only after another delay period is the last
photon from the second pair (photon 4) detected. Entanglement
swapping creates correlations between the first and last photons
non-locally not only in space, but also in time. Quantum
correlations are only observed a posteriori, after the measurement
of all photons is completed.

\begin{figure}[tb]
\centering\includegraphics[angle=0,width=86mm]{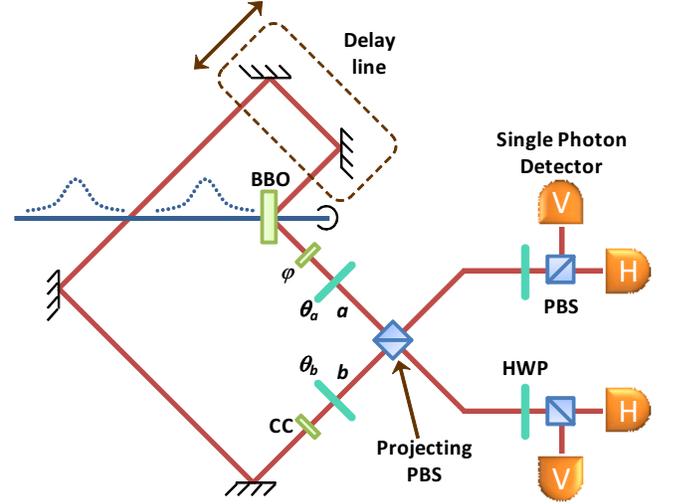}
\caption{\label{ExperimentalSetup}(color online). The experimental
setup (see text for details).}
\end{figure}

We realized this scenario with the experimental setup presented in
Fig. \ref{ExperimentalSetup} \cite{Megidish}. Polarization
entangled photon pairs are created by non-collinear type-II PDC
\cite{Kwiat95}. A pulsed Ti:Sapphire laser source with a $76\,$MHz
repetition rate is frequency doubled to a wavelength of $390\,$nm
and an average power of $400\,$mW. The laser beam is corrected for
astigmatism and focused on a $2\,$mm thick $\beta$-BaB$_2$O$_4$
(BBO) crystal. Half wave plates (HWP) and compensating crystals
(CC) correct for temporal walk-offs. Tilting of the compensating
crystal in path $a$ is used to control the phase $\varphi$ of the
state, e.g., for $\varphi= \pi$ the resulting state is
$|\psi^{-}\rangle$ \cite{Kwiat95}. The $780$\,nm wavelength
down-converted photons are spatially filtered by coupling them
into and out of single-mode fibers, and spectrally filtered by
using $3\,$nm wide bandpass filters (not shown).

One photon from the first pair is delayed until another pump pulse
arrives at the generating crystal by a 31.6\,m ($105\,$ns)
free-space delay line. The delay is built from high reflecting
dielectric mirrors, with an overall transmittance higher than
$90\%$ after $10$ reflections. Less than $10\%$ of the signal is
sampled into a single mode fiber as a feedback signal that is used
to stabilize the delayed beam's spatial properties, by tilting a
piezo-mounted mirror in the middle of the delay line. We chose the
delay length to be the time between eight consecutive laser
pulses, in order not to lose signal due to the dead-time of the
single-photon detectors (Perkin Elmer SPCM-AQ4C), and to provide
enough time for the measurement of the first photon before the
second pair is created. The delayed photon of the first pair and
the non-delayed photon of the second pair are projected onto a
Bell state by combining them at the projecting polarizing
beam-splitter (PBS) (see Fig. \ref{ExperimentalSetup})
\cite{Weinfurter94}. We post-select the cases where each photon
exits this PBS at a different port. We ensure that the photons are
indistinguishable, i.e., no information is available as to whether
both were transmitted or both were reflected. After passing
through the PBS, the photons are rotated by HWPs to the
$|p/m\rangle=\frac{1}{\sqrt{2}}(|h\rangle\pm |v\rangle)$
polarization basis. For reasons of complementarity, we also define
the circular polarization basis
$|r\rangle=\frac{1}{\sqrt{2}}(|h\rangle+i|v\rangle)$ and
$|l\rangle=\frac{1}{\sqrt{2}}(i|h\rangle+|v\rangle)$. When the
polarizations of the middle photons are correlated ($hh$ or $vv$)
they are projected onto a $|\phi^{+}\rangle_{a,b}^{\tau,\tau}$
state. When they are anti-correlated ($hv$ or $vh$ ) they are
projected onto a $|\phi^{-}\rangle_{a,b}^{\tau,\tau}$ state.

In order to fully characterize the first and last photons' state,
a quantum state tomography (QST) procedure is required
\cite{James01}. Generally, such a procedure involves independent
polarization rotations of each of the photons involved. The
photons that we are interested in characterizing (1 and 4) are
measured by the projecting PBS. Before this PBS, they share the
same paths with the other two photons that entangle them (2 and
3). Thus, arbitrary rotations of HWP angles $\theta_a$ and
$\theta_b$ will also affect the required entangling projection and
the whole projected state. It would appear that fast polarization
rotators are required to selectively rotate some photons and not
others before the projecting PBS. We have found a way to
circumvent this by relying on the entangled nonlocal nature of the
initial photon pairs.

\begin{table}[tb]
\caption{The values of the three angles and their corresponding
projection measurements that were used for a complete QST of the
first and last photons. The first 5 settings generate projections
onto orthogonal states. The other four project onto elliptical
polarizations, where $|e_{xy}\rangle=\alpha(|x\rangle+|y\rangle)$
and $\alpha$ is the normalization factor.}\label{MeasBasisTable}
\begin{center}
\begin{tabular}{ c c c c c l }
   \hline \hline
  &         & $\theta_a$     & $\varphi$ & $\theta_b$  & Polarization states        \\
  \hline
 & 1        & $0$            & $0$          & $0$       & $|hh\rangle$, $|hv\rangle$, $|vh\rangle$, $|vv\rangle$         \\
 & 2        & $22.5^\circ$         & $0$          & $0$       & $|pp\rangle$, $|mm\rangle$, $|mp\rangle$                                   \\
 & 3        & $22.5^\circ$      & $90^\circ$ & $0$       & $|pl\rangle$, $|ml\rangle$                                                             \\
 & 4        & $22.5^\circ$      & $90^\circ$ & $-22.5^\circ$   & $|ll\rangle$, $|lr\rangle$                                                             \\
 & 5      & $0$         & $90^\circ$ & $22.5^\circ$      & $|rm\rangle$                                                                                       \\
 & 6      & $0$         & $90^\circ$ & $11.25^\circ$     & $|e_{hr}e_{hm}\rangle$                                                                             \\
 & 7      & $11.25^\circ$     & $90^\circ$ & $0$         & $|e_{hm}e_{hl}\rangle$                                                                             \\
 & 8      & $11.25^\circ$     & $90^\circ$ & $45^\circ$        & $|e_{vp}e_{vl}\rangle$                                                                             \\
 & 9     & $45^\circ$        & $90^\circ$ & $11.25^\circ$      & $|e_{vr}e_{vp}\rangle$                                                                             \\
\hline \hline
\end{tabular}
\end{center}
\end{table}

The polarization of a single photon that belongs to an entangled
pair is undefined (completely mixed) until the polarization of the
other photon is measured. Furthermore, the polarization of each of
the photons depends on the specific basis in which the
polarization of the other photon is measured. As the projection of
the two middle photons is always on the $h/v$ basis, their
rotation is manifested as a nonlocal rotation of the first and
last photons. Thus, the polarization of the first photon is
affected both locally by the phase $\varphi$ and the HWP angle
$\theta_a$ and non-locally by the HWP angle $\theta_b$. Similarly,
the polarization of the last photon is affected locally by the HWP
angle $\theta_b$ and non-locally by the phase $\varphi$ and the
HWP angle $\theta_a$. The overall operations on these photons are
\begin{eqnarray}\label{Rotation1}
\hat{M}_1(\theta_a,\varphi,\theta_b)&=&\hat{R}(\theta_a)
                                \left[
                                  \begin{array}{cc}
                                    1 & 0 \\
                                    0 & e^{i\varphi} \\
                                  \end{array}
                                \right]
\sigma_{x}\hat{R}(\theta_b)\sigma_{x}\,,\\
\hat{M}_4(\theta_a,\varphi,\theta_b)&=&\hat{R}(\theta_b)\sigma_{x}
                                \left[
                                  \begin{array}{cc}
                                    1 & 0 \\
                                    0 & e^{i\varphi} \\
                                  \end{array}
                                \right]
\hat{R}(\theta_a)\sigma_{x}\,.
\end{eqnarray}
where $\hat{R}(\theta)$ is a HWP rotation, and the $\sigma_{x}$
operation is a pauli rotation due to the anti-correlation in the
initial $|\psi\rangle$ state. Even though the two rotations are
not independent, it is possible to rotate the first and last
photons differently. Nevertheless, not any arbitrary rotation is
possible due to the rotations dependence. We have found 9 angle
settings that enable the projection onto 16 independent states and
a complete QST of the first and last photons (see Table
\ref{MeasBasisTable}).

\begin{figure}[tb]
\centering\includegraphics[angle=0,width=86mm]{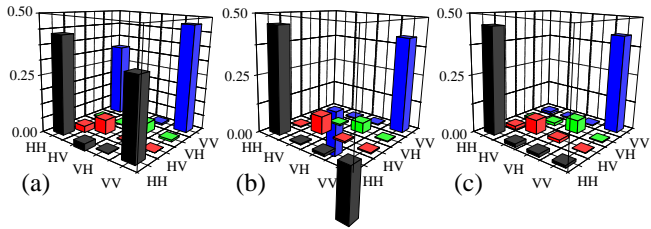}
\caption{\label{Results1}(color online). Real parts of the density
matrices of the first and last photons: (a) when the two middle
photons are projected onto the $|\phi^+\rangle$ state, (b) when
the two middle photons are projected onto the $|\phi^-\rangle$
state, and (c) when the projection fails due to temporal
distinguishability.}
\end{figure}

The density matrix of the first and last photons was constructed,
conditioned on the outcome of the projection of the two photons of
time $\tau$. If the projected photons were measured in the
$|\phi^{+}\rangle_{a,b}^{\tau,\tau}$ state, the first and last
photons were entangled in the $|\phi^{+}\rangle_{a,b}^{0,2\tau}$
state (see Fig. \ref{Results1}a). Alternatively, if the projected
photons were measured in the $|\phi^{-}\rangle_{a,b}^{\tau,\tau}$
state, the first and last photons were entangled in the
$|\phi^{-}\rangle_{a,b}^{0,2\tau}$ state (see Fig.
\ref{Results1}b). The fidelity between the measured and the
theoretical density matrices is $(77\pm1)\%$. Errors were
calculated using a bootstrapping test, assuming a Poissonian error
distribution. The total fourfold count rate was $12\,$Hz and each
polarization setting was integrated over $6$ minutes.

One can also choose to introduce distinguishability between the
two projected photons. In this case, the phase between the two
terms of the $|\phi\rangle$ projected state is undefined,
resulting in a mixture of $|\phi^{+}\rangle$ and
$|\phi^{-}\rangle$ in the projected state, and the first and last
photons do not become quantum entangled but classically
correlated. We observed this when we introduced a
sufficient temporal delay between the two projected photons (see
Fig. \ref{Results1}c). It is also evidence that the first and last
photons did not somehow share any entanglement before the
projection of the middle photons.

The fidelity of the measured entanglement is not perfect due to
several causes. It is affected in two ways by the entanglement
quality of the original photon pairs. The PDC process produces
some spectral distinguishability between the photons that reduces
the quality of pair entanglement \cite{Mosley}, which in turn,
limits the maximal quality of the swapped entanglement. In
addition, it reduces the quality of the nonlocal rotations that
are used in the QST procedure. Another cause for reduced swapping
fidelity is the presence of higher order events. We estimate this
effect to reduce the fidelity by $\sim2\%$.

The scenario we have created is very likely to occur in future
quantum repeater realizations \cite{Briegel98}. When only one
entangled photon reaches a node, it is delayed or stored in a
quantum memory until a second photon from another entangled pair
arrives. During this waiting period, the distant photon from the
first pair can already be used. Only after the arrival of a photon
from the second pair, are the two photons projected onto a Bell
state and entanglement is generated a posteriori between the other
two distant photons.

In conclusion, we have demonstrated quantum entanglement between
two photons that do not share coexistence. Although one photon is
measured even before the other is created, full quantum
correlations were observed by measuring the density matrix of the
two photons, conditioned on the result of the projecting
measurement. This is a manifestation of the non-locality of
quantum mechanics not only in space, but also in time. The
inductive nature of the setup that was used suggests that it is
possible in principle to use it to observe multiple stage
entanglement swapping.

The authors thank the Israeli Science Foundation for supporting
this work under Grant 546/10.


\begin{thebibliography}{99}
\bibitem{EPR} A. Einstein, B. Podolsky, and N. Rosen, Phys. Rev. \textbf{47}, 777 (1935).

\bibitem{Bell} J. S. Bell, Physics \textbf{1}, 195 (1964).

\bibitem{Aspect81}A. Aspect, P. Grangier, and G. Roger, Phys. Rev. Lett. \textbf{47}, 460 (1981).

\bibitem{Kwiat95}P. G. Kwiat, K. Mattle, H. Weinfurter, A. Zeilinger, A. V. Sergienko, and Y. H. Shih, Phys. Rev. Lett. \textbf{75}, 4337 (1995).

\bibitem{Weinfurter94}H. Weinfurter, Europhys. Lett. \textbf{25}, 559 (1994).

\bibitem{Vaidman99}L. Vaidman and N. Yoran, \pra \textbf{59}, 116 (1999).

\bibitem{Lutkenhaus99}N. L\"{u}tkenhaus, J. Calsamiglia, and K.-A. Suominen, \pra {\bf 59}, 3295 (1999).


\bibitem{Kim} Y.-H. Kim, S. P. Kulik, and Y. Shih, Phys. Rev. Lett. \textbf{86}, 1370 (2001) .

\bibitem{Barreiro} J. T. Barreiro, T. C. Wei, and P. G. Kwiat, Nature Physics \textbf{4}, 282 (2008).

\bibitem{Schuck} C. Schuck, G. Huber, C. Kurtsiefer, and H. Weinfurter,  Phys. Rev. Lett.  \textbf{96}, 190501 (2006)

\bibitem{Zhao05}Z. Zhao, A.-N. Zhang, Y.-A. Chen, H. Zhang, J.-F. Du, T. Yang, and J.-W. Pan, Phys. Rev. Lett. \textbf{94}, 030501 (2005).

\bibitem{Walther05}P. Walther and A. Zeilinger, \pra \textbf{72}, 010302(R) (2005).

\bibitem{Grice11} W. P. Grice, Phys. Rev. A {\bf 84}, 042331 (2011).

\bibitem{Boschi98}D. Boschi, S. Branca, F. De Martini, L. Hardy, and S. Popescu, Phys. Rev. Lett. \textbf{80}, 1121 (1998).

\bibitem{Bennett93} C. H. Bennett, G. Brassard, C. Cr\'{e}peau, R. Jozsa, A. Peres, and W. K. Wootters, Phys. Rev. Lett. {\bf 70}, 1895 (1993).

\bibitem{Zukowski93} M. \.{Z}ukowski, A. Zeilinger, M. A. Horne, and A. K. Ekert, Phys. Rev. Lett. {\bf 71}, 4287 (1993).

\bibitem{Briegel98} H.-J. Briegel, W. D\"{u}r, J. I. Cirac, and P. Zoller, Phys. Rev. Lett. {\bf 81}, 5932 (1998).

\bibitem{PanSwapping}J.-W. Pan, D. Bouwmeester, H. Weinfurter, and A. Zeilinger,   Phys. Rev. Lett. \textbf{80}, 3891 (1998).

\bibitem{PanSwapping6Photons}A. M. Goebel, C. Wagenknecht, Q. Zhang, Y.-A. Chen, K. Chen, J. Schmiedmayer, and J.-W. Pan, Phys. Rev. Lett. \textbf{101}, 080403  (2008).

\bibitem{Wiegner11}R. Wiegner, C. Thiel, J. von Zanthier, and G. S. Agarwal, Opt. Lett. \textbf{36}, 1512 (2011).

\bibitem{Peres00}A. Peres, J. Mod. Opt \textbf{47}, 139-143 (2000).


\bibitem{Ma12} X.-S. Ma, S. Zotter, J. Kofler, R. Ursin, T. Jennewein, \^{C}. Brukner, and A. Zeilinger, Nature Physics  \textbf{8}, 480 (2012).

\bibitem{Megidish} E. Megidish, T. Shacham, A. Halevy, L. Dovrat, and H. S. Eisenberg, Phys. Rev. Lett. \textbf{109}, 080504 (2012).

\bibitem{James01}D. F. V. James, P. G. Kwiat, W. J. Munro, and A. G. White, Phys. Rev. A \textbf{64}, 052312 (2001).




\bibitem{Mosley} P. J. Mosley, J. S. Lundeen, B. J. Smith, P. Wasylczyk, A. B. U'Ren, C. Silberhorn, and I. A. Walmsley,  Phys. Rev. Lett. \textbf{100}, 133601 (2008).

\end{thebibliography}
\end{document}